\documentclass[twocolumn,aps,preprintnumbers,prl,
,amsfonts,amsmath,amssymb,superscriptaddress,floatfix,10pt]{revtex4-1}

\usepackage{times}
\usepackage[dvipdfmx,final]{graphicx} 
\usepackage{amsmath}
\usepackage{amssymb}
\usepackage{color}
\usepackage{ulem}

\date{\empty}

\newcommand{\ket}[1]{|#1\rangle}
\newcommand{\bra}[1]{\langle#1|}
\newcommand{\braket}[1]{\langle#1\rangle}

\newcommand{\affA}{%
Department of Applied Physics, School of Engineering, 
The University of Tokyo,\\
7-3-1 Hongo, Bunkyo-ku, Tokyo 113-8656, Japan}

\newcommand{\affB}{%
Institute of Physics, Staudingerweg 7, Johannes Gutenberg-Universit\"{a}t Mainz, \\
55099 Mainz, Germany}

\iftrue 
	\newcommand{\eq}[1]{#1}
	\newcommand{\articletitle}[1]{ {#1}. }
\else
	\newcommand{\eq}[1]{{\it Math}}
	\newcommand{\articletitle}[1]{}
	\raggedright
\fi

\iffalse 
  
  \newcommand{\KM}[1]{{\color{blue}#1}}
  
  \usepackage{ulem}
  \newcommand{\delete}[1]{{\color{blue}#1}}
  \newcommand{\comment}[1]{{\color{blue}\scriptsize #1}}
\else
  
  \newcommand{\KM}[1]{}
  
  \newcommand{\delete}[1]{}
  \newcommand{\comment}[1]{}
\fi
\newcommand{\original}[1]{{}}
\newcommand\figA{a}
\newcommand\figB{b}
\newcommand\figC{c}
\newcommand\figD{d}
\newcommand\figE{e}
\newcommand\figF{f}
\newcommand\figG{g}
\newcommand\figH{h}
\newcommand\figI{i}

\newcommand\figrefSetup{Fig.~1\figA}

\newcommand\figurerefWavepacket{Figure~1\figB}
\newcommand\figrefWavepacket{Fig.~1\figB}

\newcommand\figrefSyncPurity{{Fig.~1\figC}}

\newcommand\figrefWaveFunction{Fig.~2}
\newcommand\figsrefWaveFunction{Figs.~2}

\newcommand\figrefQuadratures{Fig.~3}
\newcommand\figsrefQuadratures{Figs.~3}

\newcommand\figrefDensityMatrix{Fig.~3}

\newcommand\figurerefMaximumWaitingTime{Figure~1}

\newcommand\figrefPuritySM{{Fig.~S1\figA}}
\newcommand\figrefNumber{{Fig.~S1\figB}}
\newcommand\figurerefNumber{{Figure~S1\figB}}
\newcommand\figrefWaveFunctionSM{Fig.~S2}
\newcommand\figsrefWaveFunctionSM{Figs.~S2}

\newcommand\figrefQuadratureAll{Fig.~S3}
\newcommand\figrefQuadratureSM{Fig.~S4}

\newcommand\figrefSimulatedDensityMatrix{Fig.~S5}
\newcommand\figrefWigner{Fig.~S6}
\newcommand\figrefWignerOutputXX{{Fig.~S6\figC}}

\newcommand{\science}{Science }
\newcommand{\nature}{Nature (London) }

\newcommand{\natphys}{Nat.\ Phys.\ }

\newcommand{\prx}{Phys.\ Rev.\ X }

\newcommand{\etal}{\textit{et al.} }

\begin{document}

\title{Synchronization of optical photons for quantum information processing}

\author{Kenzo Makino}
\affiliation{\affA}
\author{Yosuke Hashimoto}
\affiliation{\affA}
\author{Jun-ichi Yoshikawa}
\email{yoshikawa@ap.t.u-tokyo.ac.jp}
\affiliation{\affA}
\author{Hideaki~Ohdan}
\affiliation{\affA}
\author{Takeshi Toyama}
\affiliation{\affA}
\author{Peter van Loock}
\affiliation{\affB}
\author{Akira Furusawa}
\email{akiraf@ap.t.u-tokyo.ac.jp}
\affiliation{\affA}


\date{\today}

\begin{abstract} 
A fundamental element of quantum information processing with photonic qubits is the nonclassical quantum interference between two photons when they bunch together via the Hong-Ou-Mandel (HOM) effect. Ultimately, many such pure photons must be processed in complex interferometric networks, and for this it is essential to synchronize the arrival times of the flying photons preserving their purity. Here we demonstrate for the first time the HOM interference of two heralded, pure optical photons synchronized through two independent quantum memories.
Controlled storage times up to 1.8 $\mu$s for about 90 events per second were achieved with purities sufficiently high for a negative Wigner function confirmed with homodyne measurements.
\end{abstract}

\maketitle

\noindent
Optical photons are a fundamental resource to encode flying quantum bits for quantum communication and computation.
In particular, in linear-optics quantum information processing \cite{Knill2001},
universal two-qubit gates rely upon nonclassical quantum interferences, where photons tend to bunch due to their bosonic nature.
The elementary manifestation for this is the so-called Hong-Ou-Mandel (HOM) effect \cite{Hong1987}:
when two indistinguishable single photons \eq{$\ket{1,1}$} enter a balanced beam splitter,
they bunch in either of the two output ports,
resulting in a HOM state \eq{$\ket{\text{HOM}(\theta)} = (\ket{2, 0}-e^{2i\theta}\ket{0, 2})/\sqrt{2}$} with some
relative phase \eq{$\theta$}.
For large-scale quantum computation, many pure single photons must be available simultaneously at the input ports
of large interferometric networks, in order to apply the corresponding gate sequences at the same time on all the initial qubits.
Numerous tests of the HOM effect have been performed over the past few decades mainly in order to characterize single-photon sources,
such as parametric down converters \cite{deRiedmatten2003}, trapped single neutral atoms \cite{Beugnon2006},
ions \cite{Maunz2007}, atomic ensembles \cite{Chaneliere2007}, quantum dots \cite{Flagg2010},
and nitrogen vacancy centers in diamond \cite{Bernien2012}.
However, the simultaneous occurrence of two single photons at a beam splitter has depended on a random coincidence between two independent statistical sources. 
One possible way toward scalability is to combine statistical photon sources with quantum memories,
by which a photon is stored until the other photon is available too \cite{Felinto2006, Yuan2007}.
However, such quantum storage must not be at the expense of the single-photon purity.
Only in a very recent experiment, optical single-photon states sufficiently pure to show a negative dip in their Wigner functions \cite{Lvovsky2001}
were released in a controlled fashion (quasi on demand) from a memory system based on optical cavities \cite{Yoshikawa2013, Yoshikawa2015} or atomic ensembles \cite{Bimbard2014}.

Here we report the next significant step beyond this:
the controlled HOM interference of two nearly pure photons that emerge from two independent quantum memories employing the cavity-storage method \cite{Yoshikawa2013, Yoshikawa2015}.
The resulting states still have single-photon purities above 0.5, which is sufficient for a negative Wigner function. 
Thanks to the memories, utilizing controlled storage times of up to 1.8 $\mu$s, the output HOM state can be synchronized. 
Compared to previous works with atomic memories, the memory times of our all-optical system are of similar order \cite{Felinto2006, Yuan2007, Supplementary}, however, the purities of the 
synchronized photons are raised to an unprecedented, qualitatively different level.
We believe that this controlled, almost on-demand demonstration of the HOM effect represents a breakthrough toward scaling up photonic quantum interference experiments,
with direct applications in linear-optics quantum computation \cite{Knill2001}, quantum communication \cite{Sangouard2011}, and boson sampling \cite{Aaronson2011}.

Another important aspect of our approach, making it distinct from all postselective schemes based on particle-like click-by-click photon detections,
is the characterization of the resulting HOM state from a wave-like perspective using homodyne measurements of field quadrature amplitudes.
Here we will show that a characteristic pattern in the wave basis survives even after the active synchronization with quantum memories.
A fundamental feature of quantum mechanics is the wave-particle duality, 
and our demonstration looks at the famous HOM effect from a completely wave-like angle. 
This is not only of fundamental interest but also practically important for optical hybrid quantum information processing,
where both continuous wave and discrete particle properties are exploited for quantum state preparation, processing, and detection \cite{Takeda2013, Miwa2012}. 
Since the first demonstration \cite{Hong1987}, the HOM effect has been always demonstrated as a dip in the coincidence probability of photon detections at both output ports of the beam splitter.
However, the HOM dip only reflects a particle-like aspect of the HOM state, in which photonic particle bunching becomes manifest.
The more general, actual quantum nature of that prominent optical quantum state, such as quantum entanglement, 
cannot be revealed by correlation measurements in a fixed particle basis. 
The wave-basis image of the HOM state that we present here as a counterpart to the particle-basis HOM dip is a correlation pattern similar to a four-leaf clover that vanishes and reappears depending on the relative phase \eq{$\theta$} \cite{Hashimoto2014, Etesse2015}.
The phase-dependent pattern is actually sufficient for fully characterizing the HOM state and obtaining its density matrix \cite{Lvovsky2009}. 
However, the pattern is very fragile against optical losses and detection noises mostly because the homodyne detection is sensitive to vacuum fluctuations, unlike the HOM dip whose shape is in principle unchanged even for large optical losses.
Therefore, in order to observe the phase-sensitive clover pattern, highly pure single photons must be prepared simultaneously and detected with very low-noise homodyne detectors, which became possible only very recently \cite{Yoshikawa2013, Yoshikawa2015, Bimbard2014, Hashimoto2014, Etesse2015}.
\begin{figure}[b]
		\includegraphics[width = 86mm ]{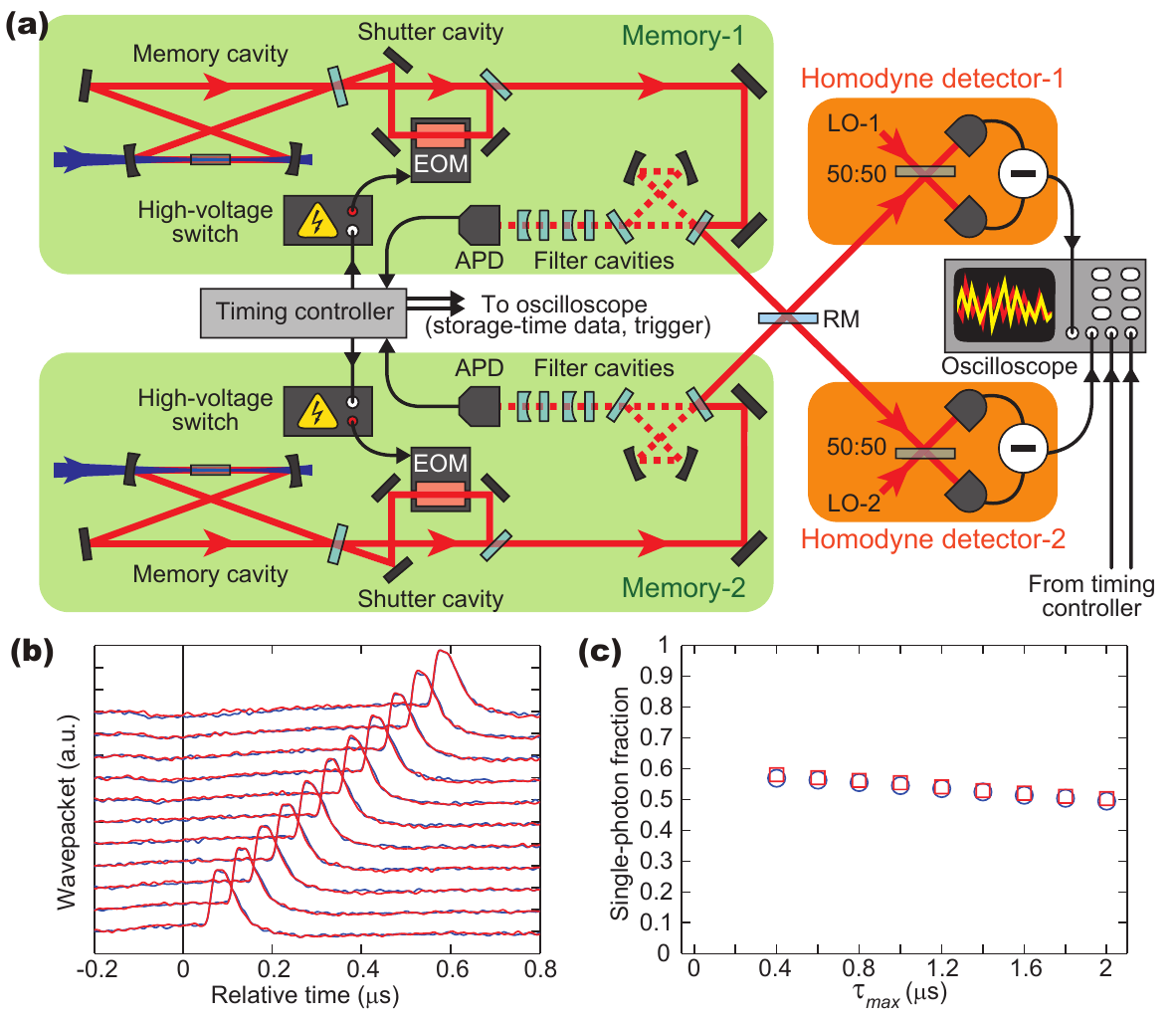}

		\caption{Experimental setup and characterization.
({\bf \figA}) Experimental setup.  
EOM, electro-optic modulator;
APD,  avalanche photodiode;
LO, local oscillator;
RM, replaceable mirror. 
({\bf \figB})  Experimentally estimated wavepackets of single photons released from Memory-1 (blue traces) and Memory-2 (red traces) after various fixed storage times.
The plots are vertically shifted according to the storage times from 0~ns to 500~ns by 50-ns steps, respectively. 
The horizontal axis is the time relative to each heralding event.
The intrinsic delay in the memory release is about 50~ns.
({\bf \figC})  Experimental single-photon purities after the synchronization without any correction of detection inefficiencies.
Circles and squares are \eq{$\bra{1}\text{Tr}_k\{\hat B^\dagger \hat\rho_\text{out} \hat B\}\ket{1}$} for Memory-1 \eq{($k=2$)} and Memory-2 \eq{($k=1$)}, respectively.
They are calculated from the output with numerical inverse of the balanced beam splitter \eq{$\hat B$}. 
}
\end{figure}

Our experimental setup is schematically shown in {\figrefSetup}, where every one of the two single-photon sources
is enclosed by a memory system (Memory-1 or Memory-2) which is composed of two concatenated cavities \cite{Yoshikawa2013, Yoshikawa2015}.
Each individual single-photon creation corresponds to an ordinary quantum optical heralding scheme,
where photons are probabilistically but simultaneously produced in pairs by nonlinear optical effects and one photon serves as the herald of the other \cite{Lvovsky2001}.
Single-photon sources based on such nonlinear optical effects have good controllability of the wavelength unlike other on-demand-type sources. 
The special interference inside the concatenated cavities ensures that the photon to be measured
is released to the outside while the photon to be prepared stays inside. 
After heralding, the stored single photon is released on demand by rapidly switching the cavity resonance
via an electro-optic effect \cite{Yoshikawa2013, Yoshikawa2015}, which is also different from a simple storage-loop switching scheme \cite{Pittman2002}.

In a preliminary experiment, we tested the performance of the individual memory systems
by using a highly reflective mirror as the replaceable mirror (RM) in {\figrefSetup}. 
For both input ports, we estimated the wavepackets of the released single photons for various fixed storage times after the heralding \cite{Yoshikawa2013, Yoshikawa2015,Morin2013}.
{\figurerefWavepacket} shows the longitudinal modes of the released wavepackets in the time domain, vertically shifted depending on the corresponding storage times. 
It can be confirmed that the single-photon wavepackets are correctly shifted by the memories without deformation,
and also that the wavepackets from the two independent memories are almost identical. 
This sameness (indistinguishability) of the wavepackets is critical for the HOM interference effect.

\begin{figure}[b]
		\includegraphics[width = 86mm]{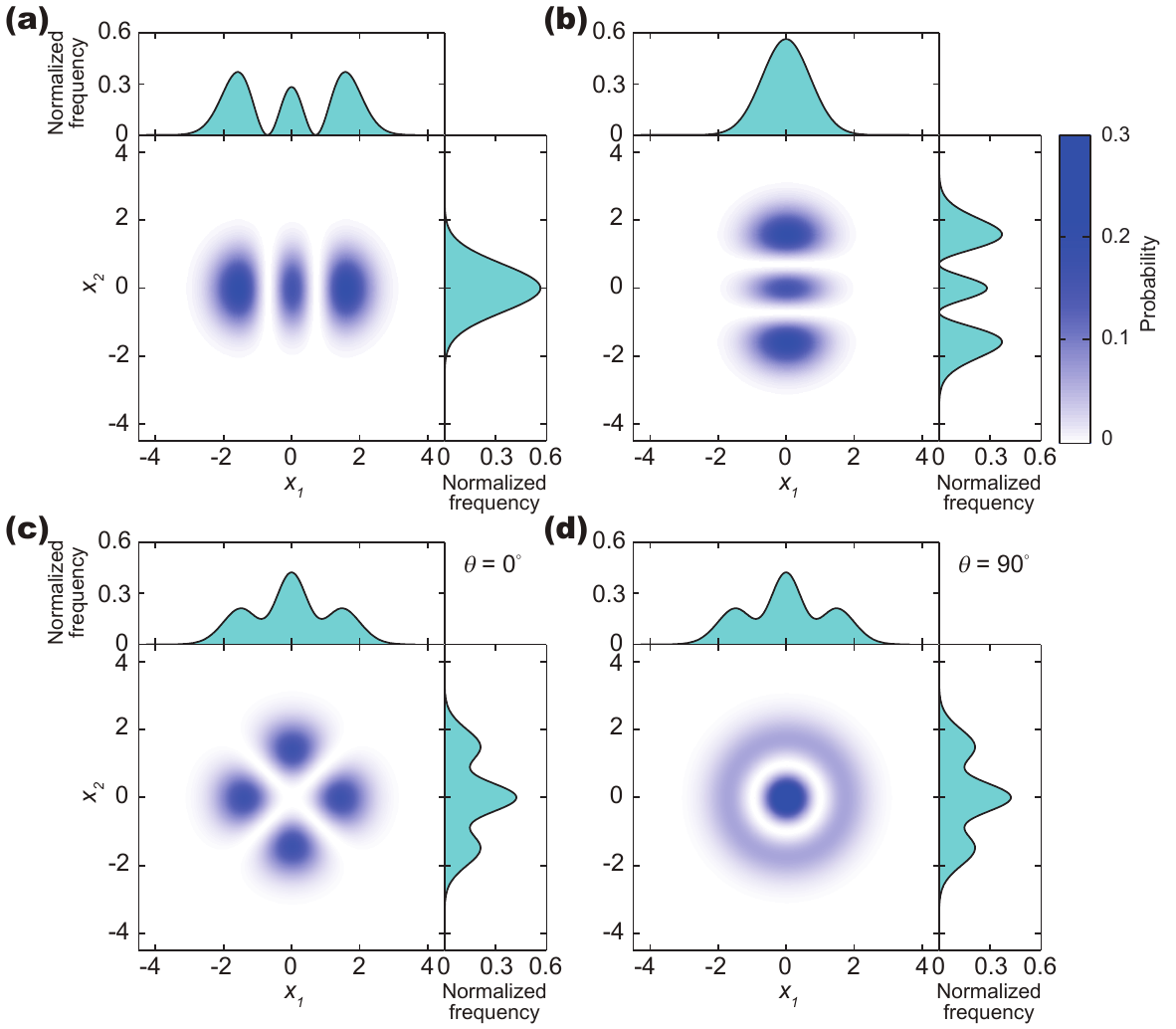}

		\caption{Theoretical two-mode quadrature distributions in the ideal cases.
({\bf \figA}) \eq{$\ket{2,0}$} state. ({\bf \figB}) \eq{$\ket{0,2}$} state.
({\bf \figC} and {\bf \figD}) Coherent superposition \eq{$\ket{\text{HOM}(\theta)}$} for $\theta = 0^\circ$ and for $\theta = 90^\circ$, respectively. 
The probability amplitudes at the origin interfere destructively (\figC) or constructively (\figD). 
The marginal distributions are also shown in the top and right panels.
		}
\end{figure}

In the HOM experiment, we synchronized the photon sources and then obtained an output HOM state.
We acquired two-mode quadrature data for 390,636 events when both single photons were heralded within 2 $\mu$s.
This large number of events was acquired within only 3 hours. 
The storage time of each event was recorded together with the homodyne outcomes. 
Then, in order to observe the dependence on the maximum storage time \eq{$\tau_\text{max}$},
we analyzed those events associated with a storage time between 0 and \eq{$\tau_\text{max}$}.
{\figurerefMaximumWaitingTime\figC} shows the individual input single-photon purities,
calculated with the maximum-likelihood method \cite{Lvovsky2009} for each \eq{$\tau_\text{max}$}.
Although the single-photon purities degrade for longer storage times due to a finite memory lifetime \cite{Supplementary}, 
the purities were kept above the 0.5-bound of the Wigner-function-negativity condition for up to about 1.8~{$\mu$}s,
which corresponds to about 90 events per second. 
The coherence time of the wavepackets for the 0.5 criteria without synchronization is estimated at 72 ns \cite{Supplementary},
and thus the memory enhancement in the event rate can be calculated as a factor of 25.

Let us now discuss the image of the HOM state in the wave basis, as shown in {\figrefWaveFunction}.
Theoretically, vacuum \eq{$\ket{0}$}, single-photon \eq{$\ket{1}$}, and two-photon \eq{$\ket{2}$} states have one, two, and three peaks in their quadrature distributions, respectively, 
and thus the two-dimensional distributions of the separable states \eq{$\ket{2,0}$} and \eq{$\ket{0,2}$} look like those in \figrefWaveFunction\figA\ and \figrefWaveFunction\figB, respectively. 
When \eq{$\ket{2,0}$} and \eq{$\ket{0,2}$} are coherently superimposed to obtain the HOM state, \eq{$\ket{\text{HOM}(\theta)} = (\ket{2,0}-e^{2i\theta}\ket{0,2})/\sqrt{2}$},
the probability amplitudes at the origin destructively or constructively interfere depending on \eq{$\theta$} \cite{Hashimoto2014,Etesse2015}.
As a result, the remaining distribution patterns are the four-leaf clover for \eq{$\theta=0^\circ$} (\figrefWaveFunction\figC)
and a concentric pattern for \eq{$\theta=90^\circ$} (\figrefWaveFunction\figD).

%
%
These characteristic distribution patterns are indeed observed in our experiment by homodyne detections.
In {\figsrefQuadratures\figA,\figB}, the distributions of the output for \eq{$\theta = 0^\circ$} and \eq{$90^\circ$} are shown,
together with the individual histograms of the single-mode quadratures.
These are the results for a \eq{$\tau_\text{max}$} of 400 ns as a typical example.
The output barely exhibits side fringes reflecting a large two-photon component (top and right panels in {\figsrefQuadratures\figA,\figB}).
The two-mode distributions of the output completely change depending on \eq{$\theta$}.
The four-leaf clover is most pronounced at \eq{$\theta = 0^\circ$} whereas the clover totally disappears at \eq{$\theta = 90^\circ$} 
as expected from {\figsrefWaveFunction\figC,\figD} \cite{Hashimoto2014,Etesse2015}.
These distributions are totally different from that of the case without synchronization (\figrefQuadratures\figC).
The results here are only particular examples, and the full results are presented in the supplementary information \cite{Supplementary}, 
together with those for the phase-independent input dual-single-photon state.

So far, we have discussed the relative phase \eq{$\theta$} of the HOM state \eq{$\ket{\text{HOM}(\theta)} = \hat U(\theta)\ket{\text{HOM}(0)}$},
where \eq{$\hat U(\theta)=e^{i\theta \hat a^\dagger_2 \hat a_2}$} is the phase-shift operator acting on the second mode.
From a different viewpoint, this can be reinterpreted as a phase shift of the measured quadratures \eq{$\bra{x_1,x_2}\hat U(\theta)$} 
for the fixed state \eq{$\ket{\text{HOM}(0)}$}, since \eq{$\text{Pr}(x_1,x_2) = \big|\bra{x_1,x_2}\hat U(\theta)\ket{\text{HOM}(0)}\big|^2$}.
In the latter interpretation, the wave-basis two-mode distributions with various measurement phases contain the complete information 
for estimating the quantum state \cite{Lvovsky2009}.
In {\figrefDensityMatrix\figD}, the density matrix of the output state \eq{$\hat \rho_\text{out}$} is shown in the number basis. 
The ideal density operator of the HOM state is \eq{$\hat\rho_\text{HOM} = \ket{\text{HOM}(0)}\bra{\text{HOM}(0)} =\frac12(\ket{0, 2}\bra{0, 2}+\ket{2, 0}\bra{2, 0} -\ket{0, 2}\bra{2, 0} -\ket{2, 0}\bra{0, 2})$}, and this structure also becomes

\onecolumngrid	

\begin{figure}[b]
		\includegraphics[height = 100mm]{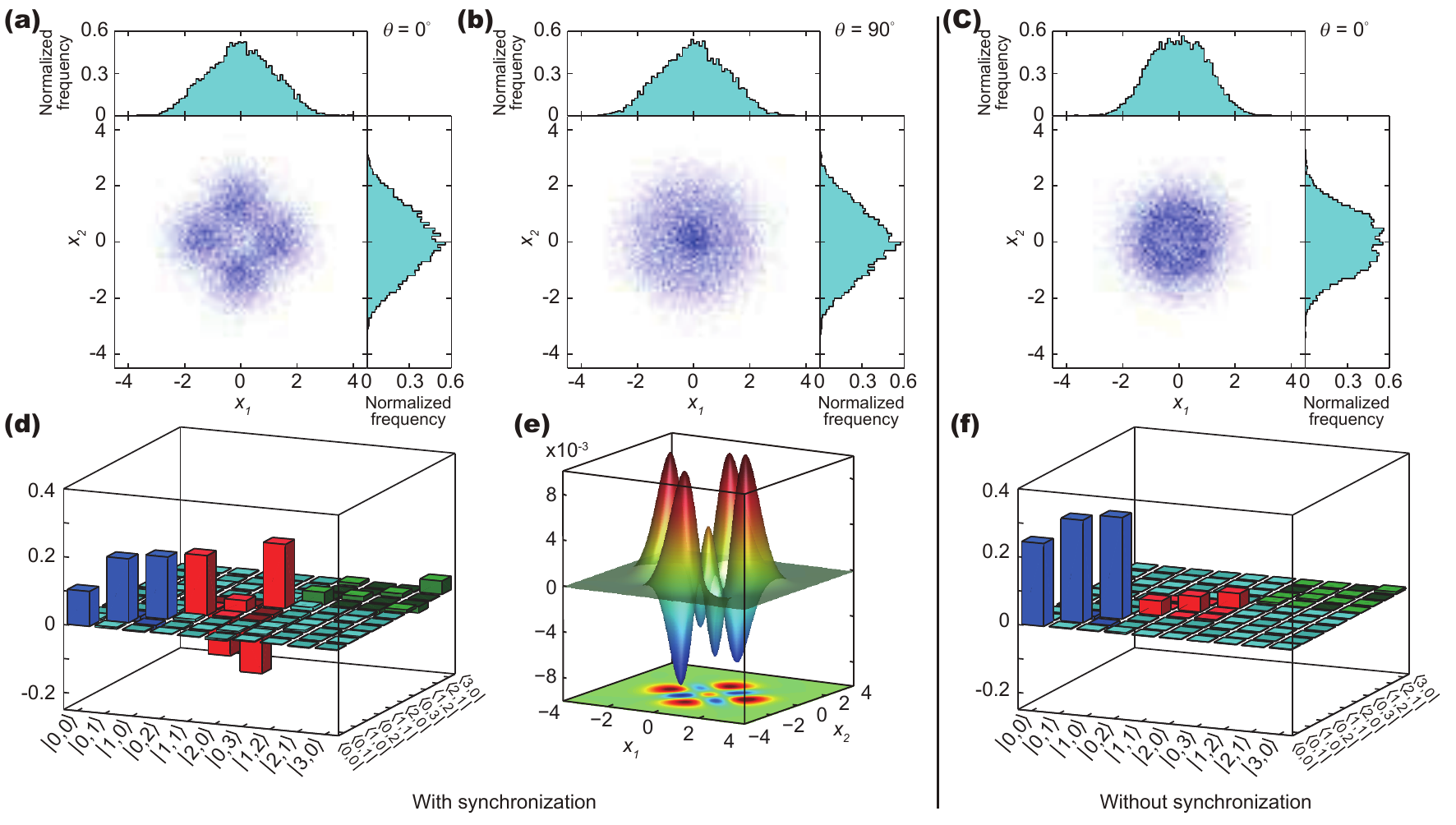}

		\caption{Experimental outputs.
({\bf \figA} and {\bf \figB}) Experimental quadrature distributions of the output state after the synchronization for \eq{$\theta = 0^\circ$} and for \eq{$\theta = 90^\circ$}.
({\bf C}) Experimental distribution without synchronization for \eq{$\theta = 0^\circ$}.
({\bf \figD}) Density matrix of the output state after synchronization. 
The imaginary part of the density matrix is omitted because it is negligibly small; it can be found in the supplementary materials \cite{Supplementary}. 
({\bf \figE}) Cross section of the output Wigner function.
({\bf \figF}) Density matrix of the output without synchronization.
}
\end{figure}

\newpage
\twocolumngrid	

\noindent
 apparent in the experimental output state \eq{$\hat\rho_\text{out}$}:
the diagonal elements of \eq{$\ket{2, 0}\bra{2, 0}$} and \eq{$\ket{0, 2}\bra{0, 2}$} indicate   the photon bunching effect on the \eq{$\ket{1, 1}\bra{1, 1}$}-component 
of the input \cite{Supplementary}, while the presence of the off-diagonal, negative elements of \eq{$\ket{0, 2}\bra{2, 0}$} and \eq{$\ket{2, 0}\bra{0, 2}$}
proves that the bunched components are quantum-mechanically superimposed rather than classically, incoherently mixed.
These off-diagonal elements are (necessary and) sufficient for the entanglement between the two output modes.
Quantitatively \cite{Plenio2005}, a logarithmic negativity \eq{$\log_2 ||\hat\rho_\text{out}{}^{T_1}|| = 0.37\pm0.01$} is obtained,
while \eq{$\log_2 ||\hat\rho_\text{HOM}{}^{T_1}|| = 1$} in the ideal case. 
Moreover, strong nonclassicality of the output state is evident from its negative Wigner function \eq{$W(x_1,p_1,x_2,p_2)$}.
The cross section \eq{$W(x_1,0,x_2,0)$ shown in \figrefDensityMatrix\figE} has negative values in a structure reflecting the four-leaf-clover pattern. A further evaluation can be found in the supplementary information \cite{Supplementary}.

In conclusion, we have experimentally demonstrated the famous HOM interference effect
in a way that is potentially scalable for large-scale quantum information processing.
To achieve this, two nearly pure single photons enter a beam splitter almost on demand
after their synchronized release from two independent optical (cavity-based) quantum memories.
We have also shown the wave-like aspect of the resulting HOM state via homodyne detections, characterized by a phase-dependent appearance of a four-leaf clover pattern. 
The quality of the states is sufficient to exhibit a negative Wigner function and entanglement even after the memory storage.
The HOM state interpreted as a NOON state \cite{Nagata2007} is also potentially useful in quantum metrology and sensing.

\vspace{1\baselineskip}

\paragraph*{Acknowledgments:}
This work was partly supported by GIA, 
PDIS, APSA of the MEXT of Japan, REFOST of Japan, the SCOPE program of the MIC of Japan.
P.v.L.\ was supported in Germany by Qcom (BMBF) and HIPERCOM (ERA-Net CHIST-ERA).
K.M.\ acknowledges support from JSPS.

\newpage
\onecolumngrid	

\begin{center}
{\bf \Large
Supplementary information for \\
synchronization of optical photons for quantum information processing
}\\

\vspace{\baselineskip}
{Kenzo Makino,$^1$ Yosuke Hashimoto,$^1$ Jun-ichi Yoshikawa,$^{1\ast}$, \\Hideaki Ohdan,$^1$ Takeshi Toyama,$^1$ Peter van Loock,$^{2}$ and Akira Furusawa$^{1\dagger}$}\\

{
\normalsize{$^{1}$Department of Applied Physics, School of Engineering, The University of Tokyo, }\\
\normalsize{7-3-1 Hongo, Bunkyo-ku, Tokyo 113-8656, Japan} \\
\normalsize{$^{2}$Institute of Physics, Staudingerweg 7, Johannes Gutenberg-Universit\"{a}t Mainz, }\\
\normalsize{55099 Mainz, Germany}\\
}

\end{center}
\twocolumngrid	

\makeatletter
	\renewcommand{\thefigure}{S\arabic{figure}}
	\renewcommand{\figurename}{Fig.~}
	\renewcommand{\thetable}{\arabic{table}}
	\renewcommand{\tablename}{\bf Table}
	\def\thetable{S\arabic{table}}
	\@addtoreset{table}{section}
	
	\long\def\@makecaption#1#2{
	\vskip\abovecaptionskip 
	\sbox\@tempboxa{{\bf#1.} #2}
	\ifdim \wd\@tempboxa >\hsize
	{\bf #1.} #2\par
	\global \@minipagefalse
	\fi
	\vskip\belowcaptionskip}

\makeatother

\setcounter{figure}{0}
\section{Supplementary Methods}

\subsection{Details of experimental setup}

The light source of our experiment is a continuous-wave (CW) Ti:sapphire laser operating at the wavelength of 860 nm.
In addition to the setup shown in \figrefSetup, there are two optical cavities which are omitted from \figrefSetup.
One cavity is a second-harmonic generator, which is a bow-tie-shaped cavity and contains a periodically-poled KTiOPO$_4$ (PPKTP) crystal 
as a nonlinear optical medium. 
The resulting continuous output beam at the wavelength of 430 nm is, after a frequency shift by an acousto-optic modulator, 
directed as a pump beam to each memory cavity, 
which contains a periodically-poled KTiOPO$_4$ (PPKTP) crystal as a nonlinear optical medium and works as 
a non-degenerate optical parametric oscillator (NOPO).
The pumping power at each memory  cavity is about 3 mW.
Optionally, the pump beams can be individually blocked after a herald in order to prevent further production of photons, 
but this was not employed in this demonstration, because the probability of such unwanted events is small.
The other cavity is a mode-cleaning cavity, by which the transverse mode of the local oscillators is 
purified to a TEM$_{00}$ mode in order to maximize the interference visibility at the homodyne detections. 
The power of each local oscillator is about 18 mW.

The concatenated-cavity-based memory systems (Memory-1, Memory-2) have the same design, 
similar to our previous experiment \cite{Yoshikawa2013} but with parameters slightly different from before.
The design of the concatenated-cavity-based memory system is as follows.
The transmissivity of the coupling mirror that couples the memory cavity and the shutter cavity is 1.6 \% 
and the transmissivity of the outcoupling mirror at the exit of the shutter cavity is 24.7\%. 
The former transmissivity corresponds to a compromise between long lifetime and phase-locking stability, 
while the latter is set in accordance with the former in order to operate the system nearly at the critical damping condition \cite{Yoshikawa2015}. 
The memory cavity has a free spectrum range (FSR) of 214.1 MHz. 
Spontaneous parametric down conversion inside the memory cavity produces signal and idler photons which are separated by this FSR. 
The shutter cavity contains an electro-optic modulator (EOM), which is a RbTiOPO$_4$ (RTP) crystal with an aperture size of 4 mm $\times$ 4 mm.
It is driven by a high-voltage switch (Bergmann Messger\"ate Entwicklung KG), whose voltage is around 900 V to match the frequency shift to the FSR of the memory cavity.

%
In order to stabilize the whole setup, all of the resonant frequencies of the cavities and the phases of the local oscillators 
are each electronically controlled by using a feedback loop. 
Each controller is composed of an analog feedback controller, an error detector of locking, and a digital scanner for the error recovery. 
The error detection signals from all of the controllers are brought together by logical OR gates, and then sent to the timing controller 
in order to pause the measurement when there is an error.
The timing controller is a field-programmable gate array (FPGA) (Virtex-4, Xilinx), 
which processes the heralding signals and controls the timing of the photon release.
The same timing controller actually also controls the sequence of the switching between the feedback 
phase and the measurement phase, as explained below. 
For the analog feedback control, we monitor the optical systems by means of bright beams.
However, such bright beams, except for the local oscillator beams and the pump beams, 
represent an extra complication in our single-photon-level experiment. 
Therefore we cyclically switch the optical systems. 
One phase is the feedback phase where the bright beams are injected to the cavity systems.
The other is the measurement phase where the bright beams are blocked and the two-photon interference is tested.
The switching rate is 5 kHz, and the duty cycle is 40\% for the measurement phase.

\subsection{Timing control sequence}
As mentioned above, the FPGA controls the release timing of photons. 
The FPGA clock frequency is 100 MHz, and thus, in our system, the release timings are synchronized with 10 ns intervals.
Note that the timing jitter of the driving signals in the timing controller was negligibly small (less than 1 ns), 
compared to the width of the wavepackets of photons (about 100 ns). 
The photon arrival times of the two inputs are matched by adjusting the electric cable lengths from the timing controller to the EOMs, 
as well as the optical path lengths from the exits of the shutter cavities to the balanced beam splitter of the HOM interferometer.
The lengths of the two output arms to the data-storage oscilloscope are also matched.
These were tested in the preliminary experiment. 

In the HOM experiment, the photons are released when both heralding events of Memory-1 and Memory-2 happen within 2 $\mu s$.
If the second herald does not occur by 2 $\mu s$,
the timing controller becomes idle and waits for 5 $\mu$s, which is long enough compared to the memory lifetime of about 2 $\mu$s. 
During this dead time, the heralded photon is almost lost and the memory systems are almost initialized again.

\section{Supplementary Discussion}

\subsection{Detailed analysis of the preliminary experiment}
In the preliminary experiment, we tested our memory systems by using a highly reflective mirror as the replaceable mirror (RM).
For each Memory-$k$, we repeated single-photon generation and homodyne detection 43,404 times for each variable 
set storage time $\tau$ after the herald, changed from 0 ns to 500 ns at 50-ns intervals.
From the results, we estimated the shapes of the wavepackets $f_k(t; \tau)$ by utilizing principal 
component analysis \cite{Morin2013, Yoshikawa2013, Yoshikawa2015}, as shown in \figrefWavepacket, and calculated the single-photon purities $P_k(\tau)$ (corresponding to the respective single-photon fractions), as shown in \figrefPuritySM.
The single-photon purities are shown to degrade for longer storage times indicating a finite memory lifetime,
but the Wigner-function-negativity condition \cite{Lvovsky2001,Yoshikawa2013} with purities above 0.5 is still satisfied for up to about  0.4~{$\mu$}s.
As shown in \figrefSyncPurity, the maximum storage time $\tau_\text{max}$ of the synchronization can be set much longer than this 0.4~$\mu$s, thus preserving the negativity. 

\begin{figure}[b]
		\includegraphics[width = 86mm]{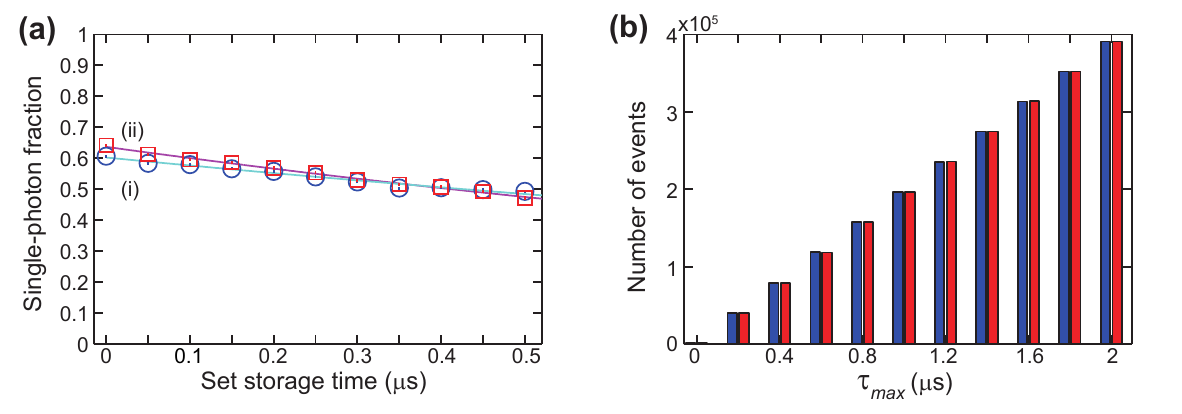} 
	
\caption{Input characterization and dependence on the maximum storage time. 
({\bf \figA}) Input single-photon purities $\bra{1}\hat{\rho}\ket{1}$ with respect to the storage time.
Blue open circles and red open squares represent Memory-1 and Memory-2. Error bars are $\pm$0.006.
Traces (i and ii) are exponential fittings.
({\bf \figB}) Number of events for the input state (blue) and the output HOM state (red)
with respect to the maximum storage time $\tau_\text{max}$.
}
\end{figure}

The shapes of the estimated wavepackets $f_k(t; \tau)$, shown in \figrefWavepacket, are independent of $\tau$, excluding pre-leakage.
Furthermore, they are almost the same between Memory-1 and Memory-2.
The overlap $C(0) =  |\int dt f_1(t;0) f_2(t;0)|^2$ was {0.992}.
We defined the signal wavepacket for the HOM experiment as an average temporal mode:
$f_\text{HOM}(t) = [f_1(t; 0) + f_2(t; 0)]/ \sqrt{\int dt | f_1(t; 0) + f_2(t; 0)|^2}$.
The overlap between the signal wavepacket and the estimated wavepacket released from the individual Memory-$k$ 
after $\tau$, defined as $C_{k}(\tau)= |\int dt  f_\text{HOM}(t-\tau) f_k(t; \tau)|^2 $, was larger than 0.96 for up to $\tau$ = 500 ns, 
as listed in Supplementary Table~\ref{Table: ModeFunction}.
This high value contributes to the visibility of the HOM interference and the purities of the input states. 

The estimated purities $P_k(\tau)$ corresponding to the estimated wavepackets $f_k(t; \tau)$ in \figrefWavepacket\
are listed in Table~\ref{Table: ModeFunction}, corresponding to the exact values of the points in \figrefPuritySM.
In order to exclude the influence of the mode mismatches on the input purities, we also calculated purities 
$P'_k(\tau)$ by using the time-shifted signal wavepacket $f_\text{HOM}(t-\tau)$, 
as also shown in Table~\ref{Table: ModeFunction}.
From the values below, we can see that the contribution of the pre-leakage is very small.
The decay of the purity is thus due to the optical losses in each memory system. 
These results are fitted with an exponential decay curve $\tilde P_k (\tau) = \tilde P_k(0) \exp(-\tau/\tilde\tau_k)$, 
where $\tilde P_k(0) $ means the initial purity and $\tilde \tau_k$ means the memory lifetime for the memory system.

\begin{table}[ht]
\begin{center}
	\caption{Overlap of the wavepackets and purity for each wavepacket.}
	\begin{tabular}{|r|cc|cc|cc|} 
	\hline
	Set storage time & \multicolumn{2}{c|}{Overlap} & \multicolumn{2}{c|}{Purity ($f_k(t; \tau)$)} & \multicolumn{2}{c|}{Purity ($f_\text{HOM}(t-\tau)$)} \\
	\hline
	$\tau$ (ns) & $C_1(\tau)$ & $C_2(\tau)$ & $P_1(\tau)$ & $P_2(\tau)$& $P'_1(\tau)$ & $P'_2(\tau)$\\ \hline 
         0 & 0.998& 0.998& 0.606& 0.642& 0.608& 0.642\\ 
        50 & 0.994& 0.992& 0.584& 0.613& 0.584& 0.613\\ 
       100 & 0.994& 0.987& 0.579& 0.596& 0.577& 0.593\\ 
       150 & 0.988& 0.986& 0.566& 0.586& 0.561& 0.581\\ 
       200 & 0.989& 0.984& 0.558& 0.569& 0.555& 0.563\\ 
       250 & 0.990& 0.982& 0.540& 0.553& 0.536& 0.546\\ 
       300 & 0.989& 0.987& 0.524& 0.529& 0.524& 0.526\\ 
       350 & 0.984& 0.984& 0.504& 0.516& 0.502& 0.515\\ 
       400 & 0.984& 0.978& 0.505& 0.507& 0.501& 0.504\\ 
       450 & 0.977& 0.978& 0.499& 0.493& 0.494& 0.487\\ 
       500 & 0.962& 0.969& 0.493& 0.472& 0.483& 0.461\\ 
      \hline
      Fitting parameters & \multicolumn{2}{c|}{$\tilde P_k(0)$} & 0.602 & 0.637 & 0.603 & 0.636 \\
      $\tilde P_k(0)\exp(-\tau/\tilde\tau)$ & \multicolumn{2}{c|}{$\tilde \tau_k$ ($\mu s$)} & 2.3 & 1.7 & 2.2 & 1.6 \\
      \hline
\end{tabular}
	\label{Table: ModeFunction}
\end{center}
\end{table}

\subsection{Dual-heralding event rate}
In the HOM experiment, we synchronized the photon sources, created an initial dual-single-photon state,
and then obtained an output HOM state by replacing the highly-reflective RM in {\figrefSetup} with a balanced beam splitter. 
The storage time of each event was recorded together with the homodyne outcomes.
Then, in order to observe the dependence on $\tau_\text{max}$,
we analyzed those events associated with a storage time between 0 and $\tau_\text{max}$.
{\figurerefNumber} shows the obtained number of events with respect to each $\tau_\text{max}$.
It indicates a linear increase of the dual-heralding event rate for any extension of $\tau_\text{max}$.

Here we discuss how the maximum storage time $\tau_\text{max}$ is dependent on the dual-heralding event rate.
We assume that the individual heralding events of the Memories occur at random and independently, i.e.\ they follow the Poisson distributions 
with average event rates $R_k$ for each Memory-$k$ ($k = 1, 2$).
Considering the (dimensionless) measurement duty $\delta$ and ignoring the 5-$\mu$s dead time for simplicity, 
the theoretical dual-heralding event rate is given by $R_\text{dual}(\tau_\text{max}) = 2\tau_\text{max} R_1 R_2/\delta$, 
when $\tau_\text{max}$ is much smaller than $\delta/R_1$ and $\delta/R_2$. 
Here the factor of 2 effectively stems from the fact that the first heralding event may occur in either of Memory-1 and Memory-2. 
This linear relationship of the dual-heralding event rate for $\tau_\text{max}$ can be confirmed in \figrefNumber.
The experimental parameters were $R_1, R_2 = 3,200$ counts per second (cps), 
and $\delta=40\%$. 
From above, the predicted dual-heralding event rate for $\tau_\text{max}=$ 1.8~$\mu s$ is $R_\text{dual}(1.8~\mu s) = 92$~cps,
which agrees well with the experimental value of about 90~cps.

\subsection{Coherence time of wavepackets}
In order to estimate the enhancement by the memories in the dual-heralding event rate, 
here we discuss the coherence time of the wavepacket without synchronization, 
from the aspect of the 0.5-bound of the Wigner-function-negativity condition. 
However, we note that this is a problem dependent on how the signal wavepacket is defined. 
A reasonable definition of the signal wavepacket is 
to set the temporal origin of the signal wavepaket to the mean of the two heralding event timings: 
when individual heralding events happen at $t_1=0$ and $t_2=\Delta t$, 
then a signal wavepacket $f_\text{HOM}(t-\Delta t/2)$ 
has a balanced overlap with both of $f_1(t;0)$ and $f_2(t-\Delta t;0)$. 
By using this definition, we calculate the average purity 
for various timing-mismatched events $(-\tau_\text{coh} \le \Delta t \le \tau_\text{coh})$ 
as a function of the maximum time difference $\tau_\text{coh}$. 
Then we find that the average purity crosses the 0.5-bound of the Wigner-function-negativity condition 
at $\tau_\text{coh}$ = 72~ns.
Comparing this $\tau_\text{coh}$ of 72~ns with the $\tau_\text{max}$ of 1.8~$\mu s$ for the synchronized case, 
and taking into account the linear dependence on $\tau_\text{max}$, 
the memory enhancement in the event rate can be estimated at a factor of 25. 

\subsection{Two-mode wave functions and probability distributions}
Let us now discuss the image of the HOM state in the wave basis by considering wavefunctions (probability amplitudes) in {\figrefWaveFunctionSM}.
Theoretically, vacuum \eq{$\ket{0}$}, single-photon \eq{$\ket{1}$}, and two-photon \eq{$\ket{2}$} states have zero, one, and two nodes in their Schr\"odinger standing wave functions \eq{$\psi_n(x)=\braket{x|n}$} 
for the field quadrature amplitude \eq{$x$}, respectively ({\figrefWaveFunctionSM\figA}).
The two-mode wave functions of separable states \eq{$\ket{2,0}$}, \eq{$\ket{0,2}$}, and \eq{$\ket{1,1}$} are
products of the single-mode wave functions ({\figsrefWaveFunctionSM\figB--\figD}).
When \eq{$\ket{2,0}$} and \eq{$\ket{0,2}$} are coherently superimposed to obtain the HOM state, \eq{$\ket{\text{HOM}(\theta)} = (\ket{2,0}-e^{2i\theta}\ket{0,2})/\sqrt{2}$},
the probability amplitudes at the origin destructively or constructively interfere depending on \eq{$\theta$}.
As a result, the superimposed wave function corresponds to a four-leaf clover for \eq{$\theta=0^\circ$} 
({\figrefWaveFunctionSM\figE}) and a concentric pattern for \eq{$\theta=90^\circ$} ({\figrefWaveFunctionSM\figF}).
Comparing {\figrefWaveFunctionSM\figE} with {\figrefWaveFunctionSM\figD} representing the input state \eq{$\ket{1,1}$},
we can see that the interference at the balanced beam splitter for \eq{$\theta = 0^\circ$} corresponds to a 45$^\circ$ rotation of the pattern.
The probability distributions \eq{$\text{Pr}(x_1,x_2)$} are the absolute squares of the wave functions ({\figsrefWaveFunctionSM\figG--\figH}).

\begin{figure}[t]
		\includegraphics[width = 86mm]{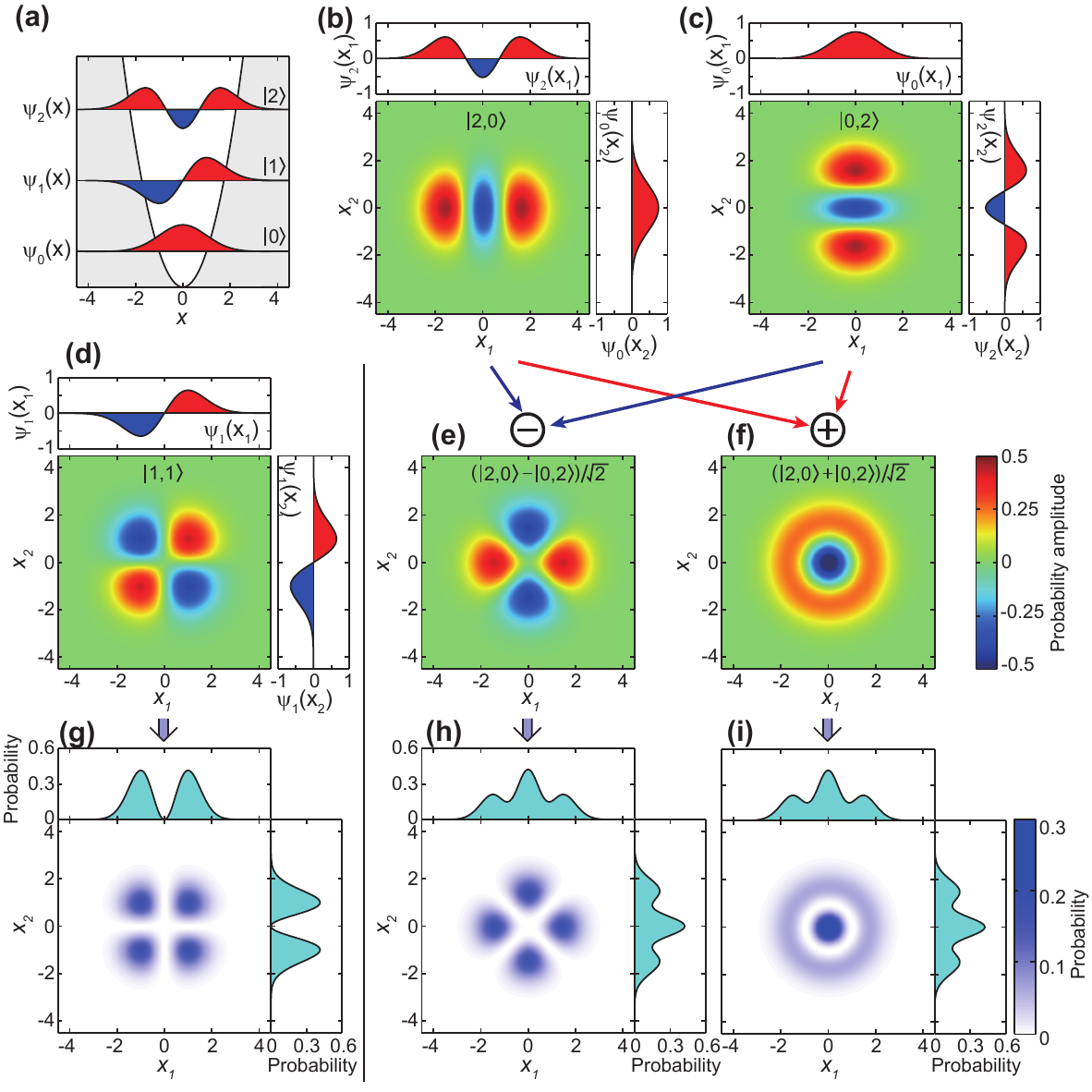} 

\caption{Wave functions and probability distributions. 
({\bf \figA}) Single-mode wave functions $\psi_n(x)$ of photon-number states.
({\bf \figB} to {\bf \figF}) Two-mode wave functions.
Those of separable states $\ket{2,0}$, $\ket{0,2}$, and input $\ket{1,1}$
are products of single-mode wave functions $\braket{x_1,x_2|n_1,n_2}=\psi_{n_1}(x_1)\psi_{n_2}(x_2)$.
At the HOM output, the probability amplitudes at the origin interfere destructively (\figE) or constructively (\figF).
({\bf \figG} to {\bf \figI}) Probability distributions $\text{Pr}(x_1,x_2)$, which are absolute squares of the wave functions ({\figD} to \figF), respectively.
The marginal distributions are also shown in the top and right panels.
}
\end{figure}

\subsection{Experimental quadrature distributions for the 36 measurement bases}
In the HOM experiment, the optical phases $(\theta_1, \theta_2)$ of the two measured quadratures $(x_1(\theta_1)$, $x_2(\theta_2))$
are independently set to $0^\circ$, $30^\circ$, $60^\circ$, $90^\circ$, $120^\circ$, and $150^\circ$.
The quadrature distributions for the tested $6\times6=36$ combinations are shown in \figrefQuadratureAll. 
These are the results for a $\tau_\text{max}$ of 400 ns as a typical example.
The distributions for equal relative phases, as can be seen in the plots along the diagonal of \figrefQuadratureAll, are almost the same, 
thus reflecting the phase-insensitivity of the input single-photon states.
The characteristic results of the output for the relative phase \eq{$\theta = \theta_1-\theta_2$} of $0^\circ$ and \eq{$90^\circ$} are collected in {\figrefQuadratureSM}, 
together with those of phase-insensitive input state.

\begin{figure*}[t]
		\hspace{-1.2cm}
		\includegraphics[width = 135mm]{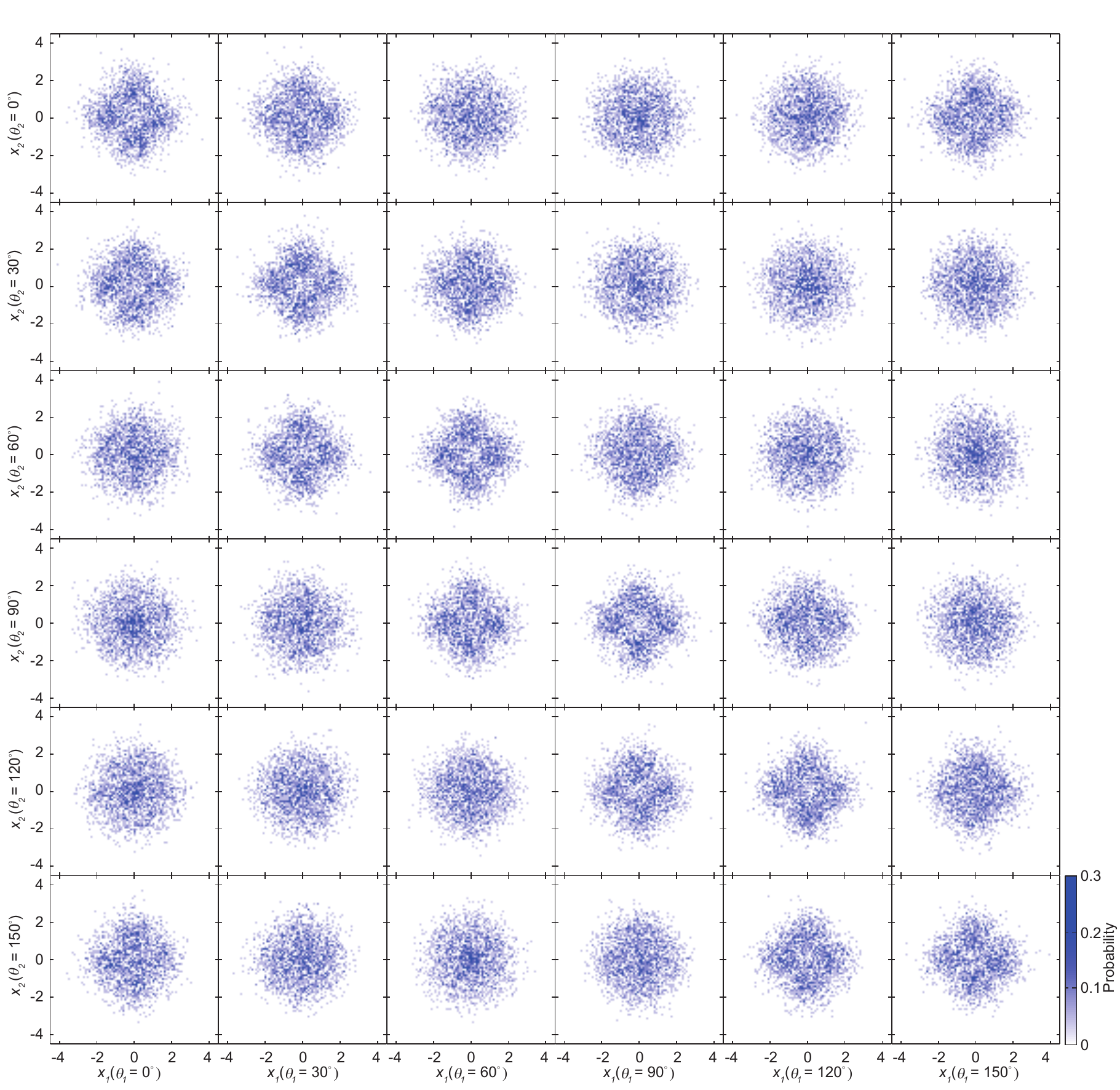} 

		\caption{Experimental two-mode quadrature distributions of the output state. 
Quadrature distributions for 36 combinations of the optical phases ($\theta_1$, $\theta_2$) are shown.
}
\end{figure*}

\begin{figure*}[t]
		\includegraphics[]{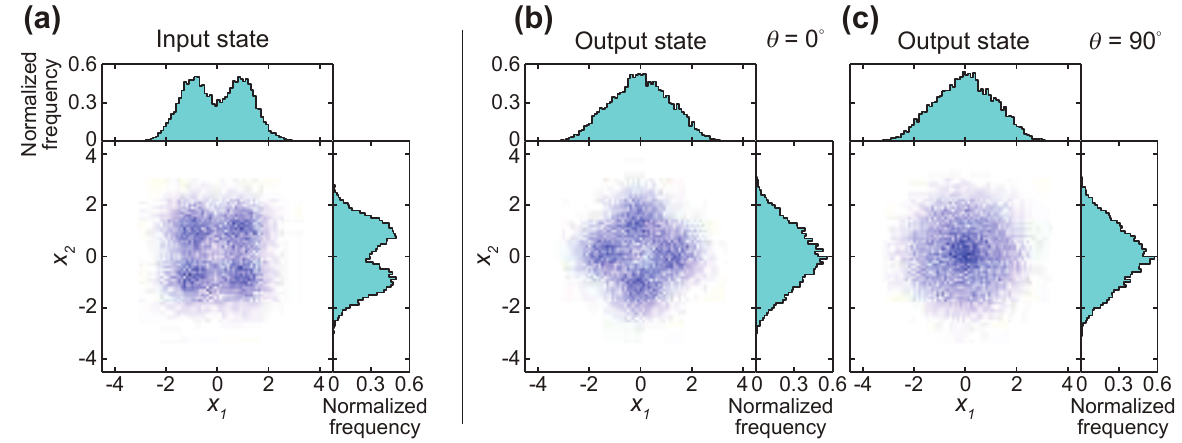} 
		
		\caption{Experimental quadrature distributions of the input and output states.
({\bf \figA} to {\bf \figC}) Experimental quadrature distributions of the input state
and the output state for \eq{$\theta = 0^\circ$} and for \eq{$\theta = 90^\circ$}.	
}
\end{figure*}

\subsection{Quantitative evaluation of the entanglement of the output HOM state}
Here we quantitatively evaluate the output state with regards to its correlations, interference visibility, and entanglement.   
The intensity cross-correlation factor, which corresponds to the depth of the HOM dip, 
$\langle   a^\dagger_1  a^\dagger_2  a_2  a_1 \rangle / \langle   a^\dagger_1  a_1 \rangle \langle   a^\dagger_2  a_2 \rangle$, is 0.24 $\pm$ 0.01. 
The interference visibility, defined as 
$\langle |  a^\dagger_1  a_1-  a^\dagger_2  a_2| \rangle / \langle   a^\dagger_1  a_1+   a^\dagger_2  a_2\rangle$
 is 0.902 $\pm$ 0.005. 
This is much higher than the phase-randomized coherent light limit of 0.5, which confirms the non-classicality of the output state.
In order to verify the quantum entanglement in the two output ports, we calculated the entanglement criterion \cite{Ray2011}.
The logarithmic negativity \cite{Plenio2005} of the output state was $\log_2 ||\hat\rho_\text{out}{}^{T_1}|| = 0.37\pm 0.01$, 
where the superscript of $T_1$ means partial transpose with respect to mode 1.
In order to exclude the contribution of undesired photon-number components to the entanglement, 
we also calculated the logarithmic negativity of the output state after numerically applying local filters $\ket{0}\bra{0}+\ket{2}\bra{2}$
to individual modes. 
The local filters do not increase the entanglement 
but just remove irrelevant photon-number components 
and transform the output state to $\rho'_\text{out}=P_\text{s}\rho_\text{s}+(1-P_\text{s})\rho^\perp$. 
Here, $\rho_\text{s}$ is the filtered density operator of $\hat\rho_\text{out}$ in the Hilbert space 
spanned by $\ket{0, 0}, \ket{0, 2}, \ket{2, 0}$, and $\ket{2, 2}$, 
$P_\text{s}$ is the probability fraction of the filtered components which was experimentally 0.470, 
and $\rho^\perp$ is a non-entangled state in the orthogonal Hilbert space. 
For this filtered output state, the logarithmic negativity was $\log_2||\rho'_\text{out}{}^{T_1}||=0.30\pm0.01$. 
From comparison of these two values, we can see the main contribution of the entanglement came from the superposition 
of $\ket{2,0}$ and $\ket{0,2}$, but there is also a small discrepancy of the log-negativities which is mainly due 
to the higher-order photon components $\ket{2, 1}$ and $\ket{1, 2}$ of the input states. 
The non-zero log-negativities of the filtered and unfiltered states confirm the entanglement of our output HOM state.

\subsection{Simulation of our HOM experiment from the input states}
By using the density matrices of the input states (\figrefSimulatedDensityMatrix~\figA,\figB), 
we theoretically calculated a density matrix of the output state with the ideal balanced beam splitter (\figrefSimulatedDensityMatrix~\figE,\figF). 
Compared to this simulation, the density matrix of the experimental output HOM state has a similar structure (\figrefSimulatedDensityMatrix~\figC,\figD).

\onecolumngrid

\begin{figure}[b]
			\includegraphics[]{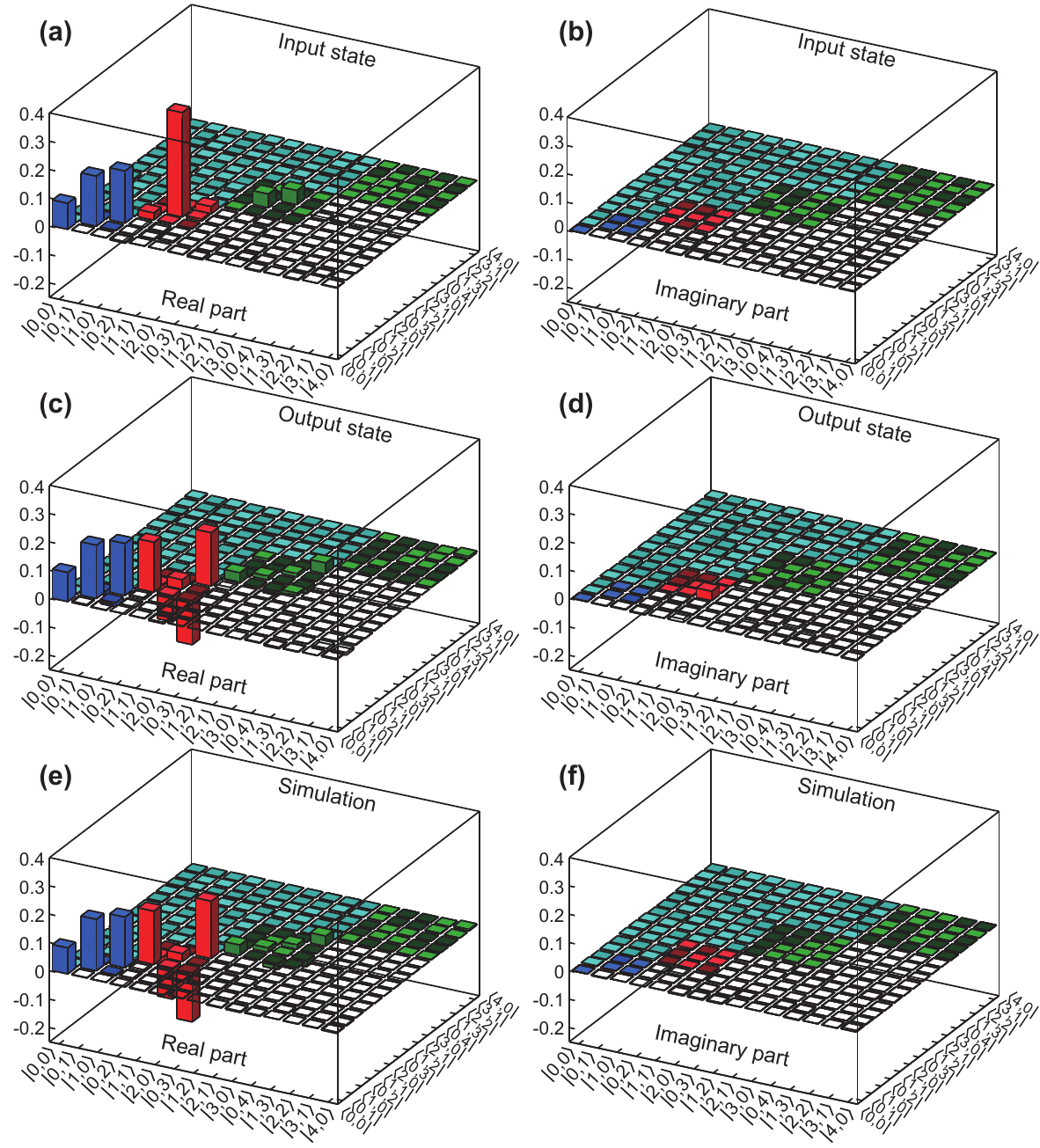} 
		
		\caption{Real and imaginary parts of the density matrices. 
({\bf \figA}, {\bf \figB}) Experimental input state.
({\bf \figC}, {\bf \figD}) Experimental output state. \newline
({\bf \figE}, {\bf \figF}) Simulated result of the output state from the experimental input state with a numerical balanced beam splitter.
}	
\end{figure}

\newpage
\twocolumngrid

\subsection{Wigner function of the output state}
Additional important information about the phase-space characteristics of the entangled HOM state can be obtained by analyzing its Wigner function. 
We calculated the Wigner function of the output state in order to verify the strong non-classicality of this state. 
The Wigner function of a two-mode state $\hat \rho$ is defined ($\hbar = 1$) in four-dimensional space as 
\onecolumngrid
\begin{equation}
	W(x_1,p_1, x_2, p_2) = \Bigl(\frac{1}{2\pi} \Bigr)^2 \iint ^\infty_{-\infty}dy_1dy_2 e^{i(p_1y_1+p_2y_2)}\Bigl\langle{x_1-\frac{y_1}{2}}, {x_2-\frac{y_2}{2}}\Bigl | \hat \rho\Bigr |{x_1+\frac{y_1}{2}}, {x_2+\frac{y_2}{2}}\Bigr\rangle \notag
	\label{Fig:Wigner}
\end{equation}
\twocolumngrid

where $x_k = x_k(\theta_k = 0), p_k = x_k(\theta_k = 90^\circ), (k = 1, 2)$ 
correspond to a pair of rotated quadratures.
In \figrefWigner, the Wigner function of the ideal HOM state and that of the experimental output state are shown, 
for different vertical scales.
As the Wigner function is very sensitive to the non-classical properties of a quantum state, 
its value quickly changes owing to experimental imperfections.  
However, the Wigner function of the output state still shows the distinctive shape of the ideal HOM state:
it has a positive value at the origin and negative values at four points around the origin in \figrefWignerOutputXX, 
which come from the strongly non-classical single-photon inputs, 
thus confirming that the output state maintains the strong non-classicality after the photon interference.   

\onecolumngrid

\begin{figure}[b]
		\hspace{-1.2cm}
			\includegraphics[width = 130mm]{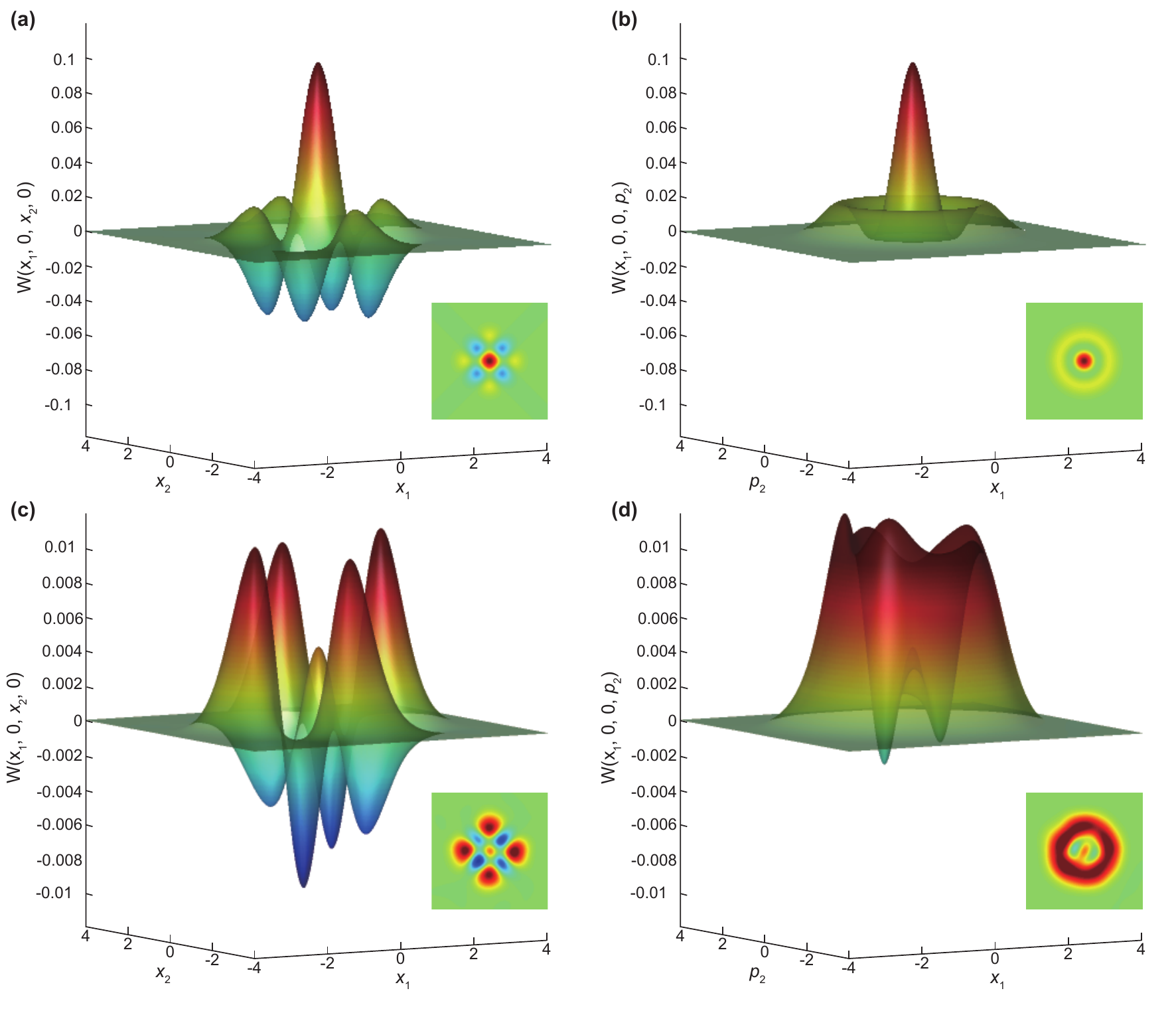} 

\caption{Cross sections of the two-mode Wigner functions. 
({\bf \figA}, {\bf \figB}) Ideal HOM state $\hat \rho_\text{HOM}$ for comparison. 
({\bf \figC}, {\bf \figD}) The experimentally reconstructed output state for $\tau_\text{max} = 400$ ns. 
Note that the vertical scales in (\figA, \figB) are different in (\figC, \figD).
The fixed cutting plane ($p_1 = 0, p_2 = 0$) is chosen to make the quadrature correlations visible (\figA, \figC).
The fixed cutting plane ($p_1 = 0, x_2 = 0$) is chosen (\figB, \figD).
}
\end{figure}

\twocolumngrid

\end{document}